\documentclass[final,1p,times]{elsarticle} 
\usepackage{graphicx} 
\usepackage{amssymb} 
\usepackage{amsthm} 
\usepackage{lineno} 

\newcommand{\ben}{\begin{eqnarray}}
\newcommand{\een}{\end{eqnarray}}

\newcommand{\bef}{\begin{figure}[h!]\centering}
\newcommand{\eef}{\end{figure}}

\journal{Nuclear Physics A} 
\begin{document} 

\begin{frontmatter} 


\title{Low-mass dilepton production in $pp$ and $AA$ collisions}

\author{Zhong-Bo Kang $^a$, Jian-Wei Qiu $^a$, and Werner Vogelsang $^b$}

\address{$^a$ Department of Physics and Astronomy, 
             Iowa State University,  
             Ames, IA 50011, USA \\
             $^b$ Physics Department, Brookhaven National Laboratory,
             Upton, NY 11973, USA}

\begin{abstract} 
We adopt a factorized QCD formalism to describe 
the transverse momentum distribution of low-mass lepton pairs
produced in $pp$ collisions, when the pair transverse momentum 
$Q_T \gg Q$, with the pair's invariant mass $Q$ as low as 
$Q \sim \Lambda_{\mathrm{QCD}}$.
We extend this formalism to dilepton production in $AA$
collisions by including the nuclear-dependent power correction due to 
parton multiple scattering.
\end{abstract} 

\end{frontmatter} 



\section{Introduction}\label{intro}

Dilepton production in ultrarelativistic heavy-ion collisions is among the richest sources of information on the features of the produced strongly-interacting medium. Since leptons do not participate in the strong interaction directly, dilepton production could probe the entire space-time evolution of the expanding hot dense system without the interference of final-state interactions. An enhanced dilepton yield from the thermal emission of the hot medium could be a signal of the deconfining phase transition for the strongly interacting matter \cite{thermal-gamma}.

The PHENIX collaboration at the Relativistic Heavy Ion Collider (RHIC) 
has recently measured the transverse momentum
distribution of electron-positron pairs, with the pairs' invariant mass as low as $100 < Q <300$ MeV, in both $pp$ and $AA$ collisions. For 
$Q_T \lesssim 3$ GeV they discovered a strong excess of dileptons in $AA$ collisions in comparison with the binary-collision scaled $pp$ result \cite{Akiba}. 

To have a reliable prediction for the dilepton yield in this low-mass 
regime $Q_T \gg Q \sim \Lambda_{\mathrm{QCD}}$ is a challenge for
perturbative QCD, even in the case of $pp$ collisions. We argue that 
the distribution at large $Q_T$ can be systematically factorized 
into universal parton-to-dilepton fragmentation functions, parton
distributions, and perturbatively calculable partonic hard parts 
evaluated at a short distance scale $\sim {\cal O}(1/Q_T)$ 
\cite{Kang:2008wv}. 
We introduce a model for the input dilepton fragmentation 
functions at a scale $\mu_0\sim 1$~GeV, which are then evolved 
perturbatively to scales relevant at RHIC. Using the evolved 
fragmentation functions, we evaluate the transverse momentum 
distributions in $pp$ collisions and are able to describe the PHENIX data in $pp$ collisions quite well.

We further extend our formalism to
dilepton production in $AA$ collisions by considering both the
leading-power nuclear effects from nuclear parton distribution functions (nPDFs), 
and the nuclear-dependent power corrections generated by parton multiple 
scattering. We find that the shadowing of nPDFs suppresses the dilepton yield while the nuclear power correction enhances it. But, the nuclear power correction alone is not sufficient to explain the large dilepton excess 
observed by PHENIX. We also explore the possibility that in $AA$ collisions 
the hot dense medium is produced before the hard collision takes place
and discuss the impact this would have on generating the dilepton excess.

\section{QCD resummation formalism for low-mass dilepton yield in $pp$ collisions}\label{pp}

The cross section for Drell-Yan type inclusive dilepton production
in hadronic collisions, 
$A(P_A)+B(P_B)\rightarrow \gamma^*(\rightarrow \ell^+\ell^-(Q))+X$, 
can be expressed in terms of the cross section for producing a
virtual photon that decays into the observed dilepton:
\begin{equation}
\frac{d\sigma_{AB\rightarrow \ell^+\ell^-(Q) X}}{dQ^2\,dQ_T^2\,dy}
= \left(\frac{\alpha_{\mathrm{em}}}{3\pi Q^2}\right)
\sqrt{1-\frac{4m_\ell^2}{Q^2}}\left(1+\frac{2m_\ell^2}{Q^2}\right)
  \frac{d\sigma_{AB\rightarrow \gamma^*(Q) X}}{dQ_T^2\,dy}\, .
\label{DY-Vph}
\end{equation}
When both physically measured momentum scales $Q$ and $Q_T$ 
are much larger than $\Lambda_{\rm QCD}$, 
the cross section for producing the virtual 
photon can be factored systematically in QCD perturbation theory as 
\cite{Collins:1989gx}
\begin{equation}
\frac{d\sigma_{AB\rightarrow \gamma^*(Q) X}}{dQ_T^2\,dy}
=\sum_{a,b}\int dx_1 f_a^A(x_1,\mu) 
           \int dx_2 f_b^B(x_2,\mu)\,
 \frac{d\hat{\sigma}^{\rm Pert}
                    _{ab\rightarrow \gamma^*(Q) X}}{dQ_T^2\,dy}
 (x_1,x_2,Q,Q_T,y;\mu) \ ,
\label{Vph-fac}
\end{equation}
where $\sum_{a,b}$ runs over all parton flavors and $d\hat{\sigma}^{\rm Pert}_{ab\rightarrow \gamma^*(Q) X}/dQ_T^2 dy$ represents corresponding perturbatively calculable short-distance hard parts. When $Q_T \gg Q$, high order hard parts could receive large corrections in powers of $\alpha_s\ln(Q_T^2/Q^2)$ caused by the radiation of partons along the direction of the low-mass dilepton. Such large logarithmic corrections can be
systematically resummed into the parton-to-virtual photon 
fragmentation functions, $D_{f\to\gamma^*}$ \cite{Berger:2001wr}.
The perturbative hard parts, 
$d\hat{\sigma}^{\rm Pert}
              _{ab\rightarrow \gamma^*(Q) X}/dQ_T^2 dy$,
can be re-organized into two terms as 
\cite{Berger:2001wr},
\begin{eqnarray}
\frac{d\hat{\sigma}^{\rm Pert}_{ab\rightarrow \gamma^*(Q) X}}
     {dQ_T^2\,dy}(x_1,x_2,Q,Q_T,y;\mu)
=
\frac{d\hat{\sigma}^{\rm Dir}_{ab\rightarrow \gamma^*(Q) X}}
     {dQ_T^2\,dy}
+
\frac{d\hat{\sigma}^{\rm Frag}_{ab\rightarrow \gamma^*(Q) X}}
     {dQ_T^2\,dy}\, , 
\label{fac-dir-frag}
\end{eqnarray}
where the superscripts ``Dir'' and ``Frag'' represent the
``direct'' and the ``fragmentation'' contribution, respectively.
The fragmentation contribution can be further factorized 
as \cite{Berger:2001wr}, 
\begin{equation}
\frac{d\hat{\sigma}^{\rm Frag}_{ab\rightarrow\gamma^*(Q) X}}
     {dQ_T^2 dy}
= \sum_{c} \int \frac{dz}{z^2}\, 
  \left[
  \frac{d\hat{\sigma}_{ab\rightarrow c X}}
       {dp_{c_T}^2\,dy}\left(x_1,x_2,p_c=\hat{Q}/z;\mu_F\right)
  \right]
  D_{c\rightarrow \gamma^*}(z,\mu_F^2;Q^2) ,
\label{DY-F}
\end{equation}
with all fragmentation logarithms resummed into 
$D_{f\to\gamma^*}$ and the perturbative 
$d\hat{\sigma}_{ab\rightarrow c X}/dp_{c_T}^2\,dy$ 
evaluated at the distance scale $1/Q_T$. The ``Dir'' term is a 
perturbative difference between the ``Pert'' and ``Frag'' terms when both
are expanded in powers of $\alpha_s$.

We have argued in Ref.~\cite{Kang:2008wv} that the factorization formalism for the dilepton yield at high $Q_T$ is also valid for lepton pairs with
invariant mass $Q\sim \Lambda_{\rm QCD}$. However, in this case 
the fragmentation functions, $D_{c\rightarrow \gamma^*}$, are no longer fully perturbative but introduce non-perturbative hadronic contributions to the dilepton yield. We proposed to account for these by adding a 
hadronic component to the fragmentation functions at a factorization scale 
$\mu_0\sim 1$~GeV \cite{Kang:2008wv}:
\begin{equation}
D_{f\to \gamma^*}(z,\mu_0^2;Q^2)
=
D^{\rm QED(0)}_{f\to\gamma^*}(z,\mu_0^2;Q^2)
+\kappa\, D_{f\to V}(z,\mu_0^2)\,
\frac{4\pi\alpha_{\mathrm{em}}}{f_V^2}\,
\left(1-\frac{Q^2}{m_{V}^2}\right)^3 \, ,
\label{full_frag}
\end{equation}
where $D^{\rm QED(0)}_{f\to\gamma^*}$ represents the leading order QED 
component and the size of the hadronic component is fixed by a fitting parameter $\kappa$. Following \cite{Gluck:1992zx}, we further assume that $V=\rho$, $f_\rho^2/4\pi=2.2$, and $D_{f\to \rho}\approx D_{f\to \pi}$. In the top left figure of 
Fig.~\ref{collider} we compare our calculation to the PHENIX dilepton data
in $pp$ collisions \cite{Akiba} 
under two assumptions: $\kappa=1$ (solid line) and $\kappa=0$ 
(dotted line). Our calculation based on the factorization formalism 
describes the data rather well. The hadronic component 
is very important when $Q_T\lesssim 3$ GeV.

\section{Nuclear modification in $AA$ collisions}\label{AA}

There could be many interesting sources of nuclear modification for dilepton production in $AA$ collisions. For the leading-power factorization formalism, one could have a universal nuclear effect from the nPDFs. We can study this effect by replacing the proton PDFs in Eq.~(\ref{Vph-fac}) by the nPDFs, $
f_i^p\to \left[Z\cdot f_i^{p/A}+(A-Z)\cdot f_i^{n/A}\right]/A$.
Using the EPS08 nPDFs \cite{EPS08}, we plot our calculated cross sections with $\kappa=1$ (dashed curves) in Fig.~\ref{collider} and compare to the
PHENIX data in Au+Au collisions \cite{Akiba}. For comparison, we also show the binary-collision scaled $pp$ results as dotted curves. The cross section with nPDFs is larger than the scaled $pp$ results when $Q_T > 3$ GeV due to the antishadowing effect and is consistent with the data. However, the shadowing effect in nPDFs leads to a suppression of the dilepton yield when $Q_T\lesssim 3$ GeV, where the data shows a very strong enhancement over the scaled $pp$ results.
\bef
\includegraphics[width=2.4in]{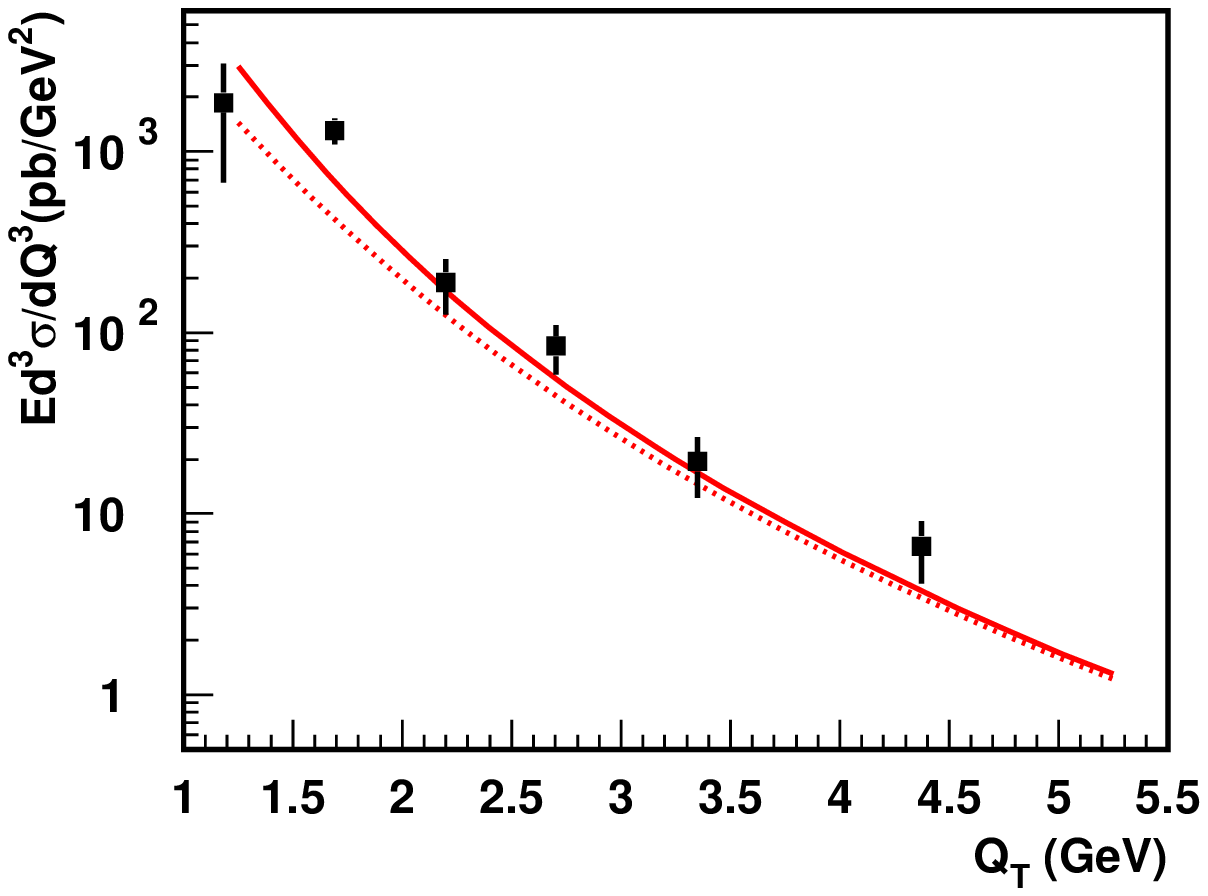}
\hfil
\includegraphics[width=2.4in]{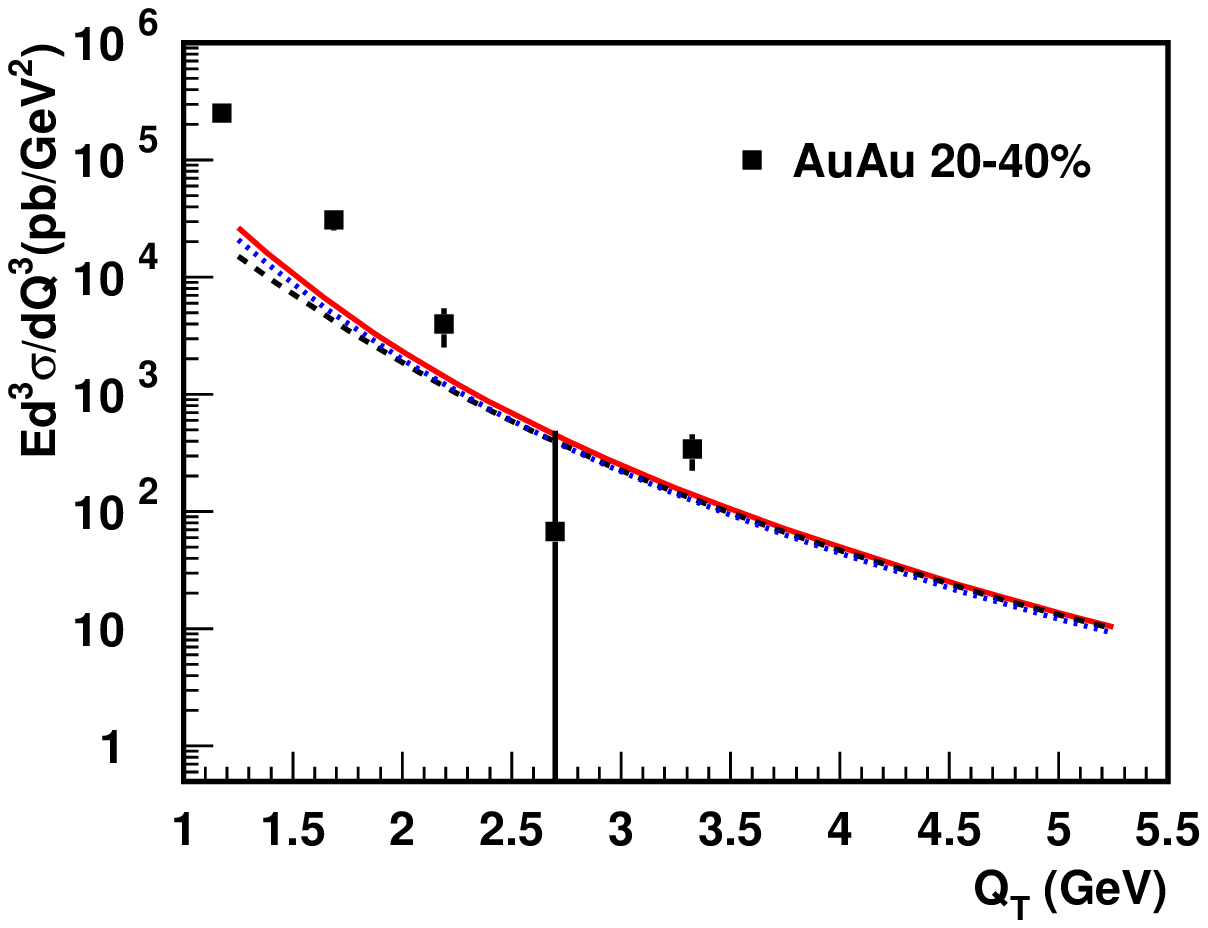}
\includegraphics[width=2.4in]{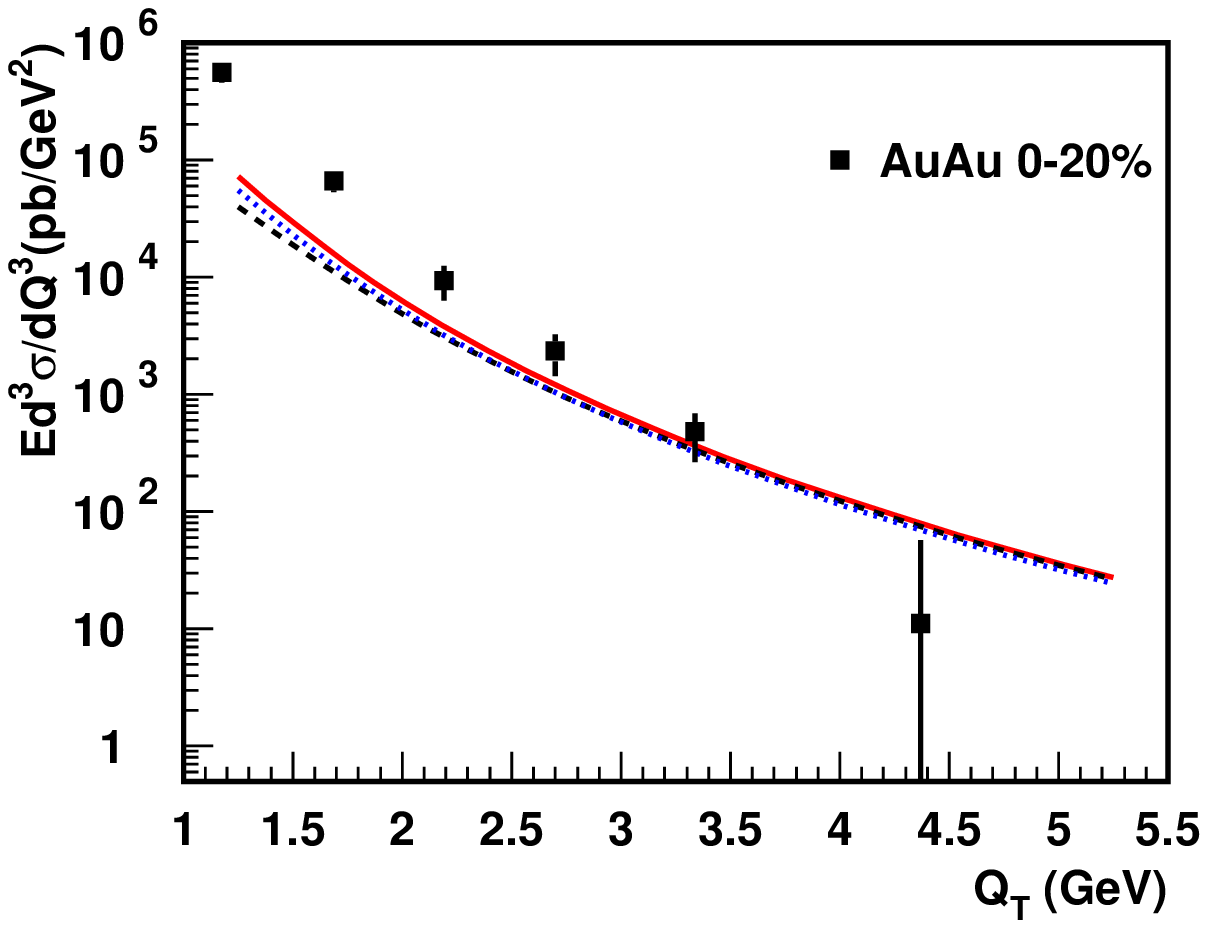}
\hfil
\includegraphics[width=2.4in]{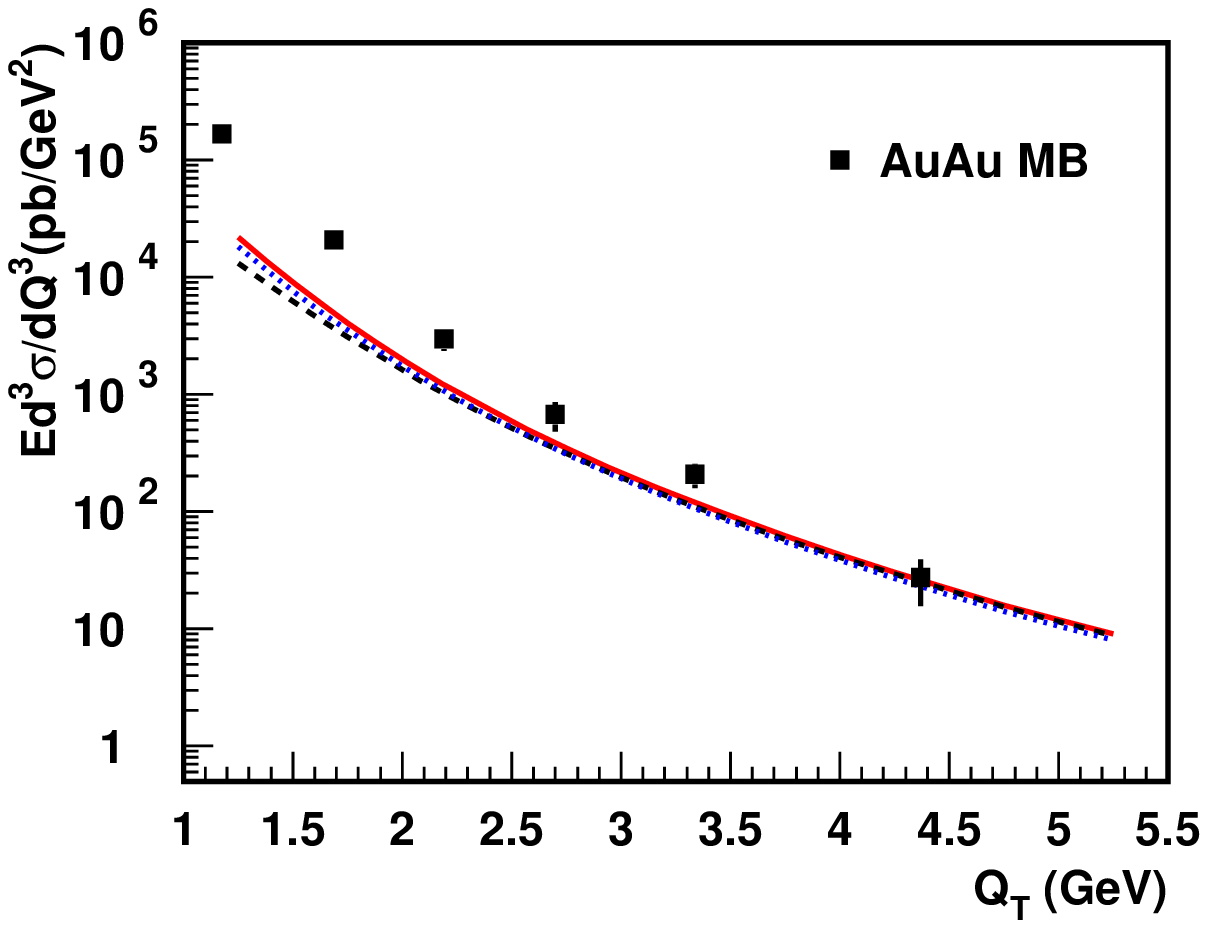}
\caption{Invariant cross section for low-mass dilepton production with $100< Q< 300$ MeV as a function of the pair's transverse momentum at $\sqrt{s_{NN}}=200$~GeV and $y=0$. Top left: 
in $pp$ collisions with $\kappa=1$ (solid) and $\kappa=0$ (dotted).
Remaining three plots: in Au+Au collisions with different centralities. The dashed and solid lines are evaluated with nPDFs, and nPDFs plus power correction, respectively. The dotted lines are for binary-collision scaled $pp$ cross section. The data are from Ref.~\cite{Akiba}.}
\label{collider}
\eef

In nuclear collisions, the high-$Q_T$ lepton pairs could also be produced by multiple scattering. When $Q$ and $Q_T$ are small, such initial-state multiple scattering could lead to medium-size enhanced power corrections to the cross section and significantly increase the yield of low-mass lepton pairs \cite{Qiu:2001zj}. The leading power correction to the dilepton yield is proportional to the twist-4 quark-gluon and four-gluon correlation functions, $T_{q/A}(x)$ and $T_{g/A}(x)$ \cite{Qiu:2001zj}. We evaluate the power correction by adopting the usual model $
T_{q,g/A}(x)=\lambda^2A^{4/3}f_{q,g}^{p/A}(x),
$
with $f_{q,g}^{p/A}(x)$ denoting as before the proton PDFs in nuclei and $\lambda^2\approx 0.01$~GeV$^2$ fixed by data \cite{Kang:2008us}. We plot our predictions 
for the dilepton yield resulting from inclusion of this power correction
as solid lines in Fig.~\ref{collider}.
We find that the power correction could enhance the yield by $70 - 90\%$ at low $Q_T$. However, the calculated cross section including the power correction is still significantly below the experimental data at low $Q_T$.

Our calculation indicates that the power correction from initial-state parton multiple scattering in cold nuclear matter cannot explain the dilepton 
excess observed by PHENIX. As argued in Ref.~\cite{Wang:2002ri}, 
the initial density of the hot medium produced in central Au+Au collisions at RHIC could be as large as 30 times that of cold nuclear matter. 
Should this hot medium (or a part of it) be formed before the hard collision that produces the high-$Q_T$ lepton pair, the rate for multiple scattering, and thus the dilepton yield from multiple scattering, could be significantly enhanced by the higher medium density. To fit the PHENIX data, we would need a hot medium density that is an order of magnitude higher than the cold nuclear density.

\section{Conclusion}

We have investigated the production cross section 
for lepton pairs in the regime $Q_T \gg Q \sim \Lambda_{\rm QCD}$ in both
$pp$ and $AA$ collisions. We have argued that the cross section for high-$Q_T$ low-mass dileptons can be treated in perturbative QCD in the same way as the cross section for direct photon production. Our calculation is consistent with the RHIC data for $pp$ collisions. We have also calculated the yield of low-mass lepton pairs in $AA$ collisions by including nuclear effects from nPDFs and parton multiple scattering in cold nuclear matter. We have found that multiple scattering enhances the dilepton yield at low $Q_T$, while the nPDFs suppress it. Although the net effect enhances the dilepton yield, the enhancement is far too small to explain the dilepton excess observed by the PHENIX collaboration. The data requires us to consider additional sources of low-mass lepton pairs in $AA$ collisions. The thermal emission of low mass lepton pairs from the hot medium could be an important source \cite{thermal-gamma}. However, because of the high transverse momentum, the thermal emission alone might not be able to generate the additional order of magnitude more lepton pairs at $Q_T > 1$~GeV needed to explain the data.

\section*{Acknowledgments} 

We thank Y. Akiba and Xin-Nian Wang for helpful discussions. This work was supported in part by the U. S. Department of Energy under Grant No. DE-FG02-87ER40371 (ZK and JQ) and contract number DE-AC02-98CH10886 (WV).

\end{document}